\documentclass{iaus}
\usepackage{graphicx}
\usepackage{upmath}
\usepackage{amsbsy}
\usepackage{amssymb}

\title[The Interplay among Black Holes, Stars and ISM in Galactic 
       Nuclei]{Ionization Mechanisms in Jet-Dominated Seyferts: A Detailed
Case Study}

\author[D. J. Rosario {\it et al.\/}]%
{D.J. Rosario$^1$, M. Whittle$^1$, J.D. Silverman$^1$, A.S. Wilson$^2$
\and C.H. Nelson$^3$}

\affiliation{$^1$Astronomy Department, University of Virginia, 
Charlottesville, VA 22903\\
$^2$Astronomy Department, University of Maryland, College Park, MD 20742\\
$^3$Physics and Astronomy, Drake University, Des Moines, IA 50311-4505}

\pubyear{2004}
\volume{222}
\jname{The Interplay among Black Holes, Stars and ISM \\in Galactic Nuclei}
\editors{Th. Storchi Bergmann, L.C. Ho \& H.R. Schmitt, eds.}
\begin{document}

\maketitle

\begin{abstract}
For the past 10 years there has been an active debate over whether
fast shocks play an important role in ionizing emission line regions
in Seyfert galaxies.  To investigate this claim, we have studied the
Seyfert 2 galaxy Mkn 78, using HST UV/optical images and spectroscopy.
Since Mkn 78 provides the archetypal jet-driven bipolar velocity
field, if shocks are important anywhere they should be important in
this object.  Having mapped the emission line fluxes and velocity field,
we first compare the ionization conditions to standard photoionization
and shock models.  We find coherent variations of ionization consistent
with photoionization model sequences which combine optically thick and
thin gas, but are inconsistent with either autoionizing shock models
or photoionization models of just optically thick gas.  Furthermore,
we find absolutely no link between the ionization of the gas and its
kinematic state, while we do find a simple decline of ionization degree
with radius. We feel this object provides the strongest case to date 
against the importance of shock related ionization in Seyferts.
\end{abstract}

\section{ Introduction }

  Ionization studies of Seyferts have a long history. Early work led 
to the establishment of nuclear photoionization as the favored NLR 
ionizing mechanism. But in the past decade or so, standard models have 
been called into question because, among other reasons, they strongly 
underestimate the strengths of many of the weaker high-ionization and 
high-excitation lines
(see \cite{binette96,robinson00} for a more complete discussion). 
This led to the development of alternative models, as well as refinements
to standard nuclear photoionization. In particular, photoionizing shocks, 
driven by AGN jets and outflows, have emerged as a viable ionizing source, 
following work by \cite{viegas89} and \cite{dopita96}. 

  We try to resolve this debate by taking the following approach: we choose 
a Seyfert with strong, NLR-wide jet-gas interactions. If shocks are important 
in providing the ionizing power in Seyferts, we should expect to see 
unambiguous signs of their presence in this object's spectrum. If not,
current refinements to nuclear photoionization can be tested.
       
\section {Mkn 78 : A Jet-Gas Interaction Archetype}

 The Seyfert 2 Mkn 78 was selected as a target because it lies well off the
virial correlation for Seyferts (\cite[Whittle 1992]{whittle92}), indicating 
the presence of widespread non-gravitational motions in the ionized gas. 
This makes it one of the best candidates for a strong radio jet/ISM 
interaction among the sample of nearby Seyferts. 
\cite{paper1} discuss the structural aspects of the interaction in detail.   

  We use a extensive dataset consisting of HST-STIS longslit spectra from 
four slits sampling all the major emission line features in the NLR
at high spatial resolution ($\sim 0.05$ arcsec). 
Our spectra give us almost complete FUV and optical wavelength coverage, 
allowing the measurement of many lines of different ionization state and
excitation level. In addition, medium resolution ($\sim 30$ km/s) spectra
allow us to accurately estimate the kinematics of the line emitting gas.  

\section{ Ionization Mechanisms }

  We consider three types of ionization models to compare to the observations.

\subsection{ Standard Nuclear Photoionization } 

  Early photoionization models 
(e.g., \cite[Davidson \& Netzer 1979]{davidson79}) invoked a population of
Lyman thick (ionization-bounded or IB) clouds illuminated 
by a power-law AGN ionizing continuum (of the form 
$F_{\nu}\propto \nu^{\alpha}$). 
Using CLOUDY (\cite[Version 94.0, Ferland 1996]{ferland96}), 
we ran a set of constant density $\alpha=-1.0$ and $\alpha=-1.4$ models,
with sequences in the ionization parameter $U=n_{i}/n_{e}=10^{-1}-10^{-3}$.
In all cases, the $\alpha=-1.4$ model was the better match to the data. 
  
\subsection{ Multi-Component Nuclear Photoionization : the \boldmath$A_{m/i}$
Sequence }         

  The $U$ models can be generalized by the introdution of matter-bounded 
or MB clouds that are optically thin to the Lyman continuum
(\cite[Binette et al. 1996]{binette96}). By varying $A_{m/i}$, the 
relative contribution of the two components to the spectrum, the range 
of observed emission line ratios can be reproduced for average Seyfert 
NLRs. Mkn 78, however, has unusually weak high-ionization lines and so 
we used CLOUDY to generate an $A_{m/i}$ sequence with the ionization 
parameter of the MB component reduced by a factor of 4 compared to the 
Binette et al. value, which allowed us to adequately match [NeV] and 
other high-ionization lines.       

\subsection{ Shock models }

  We use the \cite{dopita96} shock models, in which the hot postshock gas 
generates a photoionized quiescent ``precursor'' region. In this way, 
shock models are inherently two-component in nature, but with a 
self-consistent prescription for the relative line contribution from 
postshock and precursor material. We use models with shock velocity 
$V_{sh}=200-500\;$km/s and magnetic parameter 
$B/\sqrt{n} = 0-2\;\mu G\;\textrm{cm}^{\frac{3}{2}}$. 
We have also included a more recent sequence of equipartition 
magnetic shocks from \cite{allen04}, with $V_{sh}=200-1000$ km/s. 

\begin{figure}
 \centering
 \includegraphics[angle=-90,origin=c,scale=0.5]{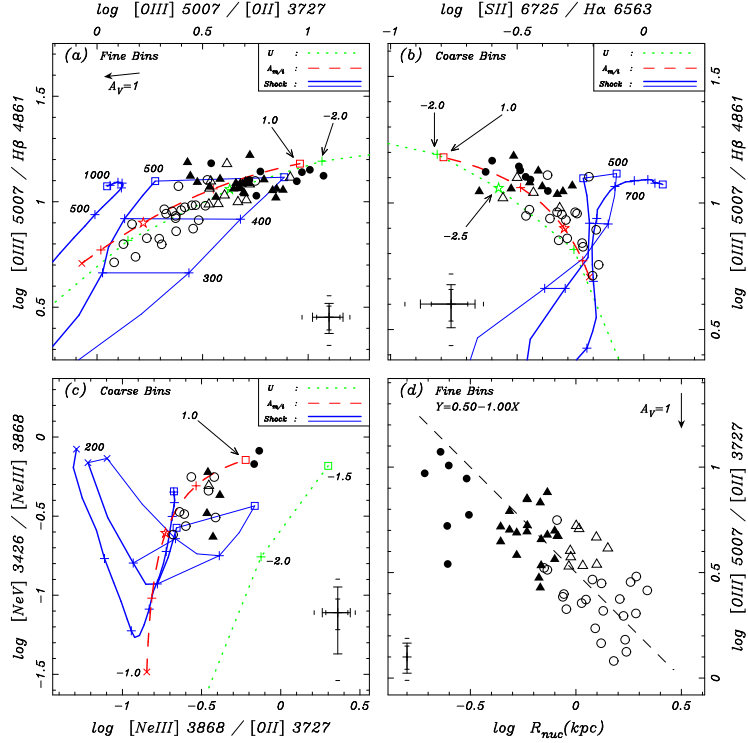}
 \caption{(a,b,c) Emission line-ratio plots. Triangles/Circles : East/West
NLR. Filled/Open : Inner/Outer NLR. Three model sequences: dotted line -- 
PL AGN photoionization, $\alpha=-1.4$, dashed line -- $A_{m/i}$, solid line --
shocks, thick/thin line: magnetic/non-magnetic shocks. All data corrected for 
extinction. Solid error bars are median $2\sigma$, with smaller ticks for 
the 10 and 90 percentile range. On the model loci, crosses are the bottom 
of the sequence ($\log U=-3.5$, $\log A_{m/i}=-1.0$, $V_{sh}=200$ km/s), 
squares are the top of the sequence ($\log U=-1.5$, $\log A_{m/i}=1.0$, 
$V_{sh}=500 \textrm{ or }1000$ km/s), plus signs are uniform steps in 
$\log U\textrm{ or }A_{m/i}=0.5$ and $V_{sh}=100$ km/s. 
(d) Excitation vs. radius The dashed line with a slope of $-1$ is not a fit. 
See \cite{paper2} for explanation of the binning scheme.}\label{fig1}
\end{figure}

\section{Methods of Analysis}

   We employ a series of tests to search for signs of shock excited gas 
and/or check the consistency of the AGN photoionization paradigm.
 
\subsection{Line-ratio vs. Line-ratio Diagrams}

   Here, we use the well-known method from \cite{baldwin81} of plotting 
sets of line-ratios vs. other line-ratios and comparing the data to the 
predictions of models. We divide these diagrams into three basic types: 

\begin{enumerate}

\item[1.] Excitation diagrams, such as [OIII]$\lambda 5007$/[OII]$\lambda 
3727$ vs. [OIII]/H$\beta$ [Fig.~\ref{fig1}(\textit{a})]. As expected, the 
data points lie in the intersection space of all the models, with implied 
$U\sim10^{-2}-10^{-3}$, $A_{m/i}\sim0.2-10.0$ and $V_{sh}\sim300-500$ 
km/s. However, trends in the data marginally support nuclear 
photoionization over shocks.

\item[2.] Shock discriminators, such as [SII]$\lambda 6720$/H$\alpha$ vs. 
[OIII]/H$\beta$ [Fig.~\ref{fig1}(\textit{b})]. The shock sequences lie 
almost perpendicular to the photoionization sequences and the trends 
in the data clearly follow the $U$ and $A_{m/i}$ model loci. 

\item[3.] U discriminators, which use line-ratios  
known to be troublesome for standard photoionization. 
Fig.~\ref{fig1}(\textit{c}) shows [NeIII]$\lambda 3868$/[OII] vs. 
[NeV]/[NeIII]. The U models predict far weaker [NeV] than is measured, 
even though this line is already unusually weak in Mkn 78. 

\end{enumerate}

 We conclude that nuclear photoionization probably dominates the 
line emission processes in the NLR of Mkn 78, with a mixture of optically 
thin and optically thick gas needed to explain the range of excitation.

\subsection{Line-ratios vs. Other Quantities}

  We can measure a host of physical and dynamical properties from our 
spectra, which we then compare to model predictions. For example, the 
\cite{dopita96} shock models predict strong correlations between shock
velocity and excitation tracing line-ratios like [OIII]/H$\beta$. 
On the other hand, the data seems to show absolutely no correlation 
between excitation and either bulk line velocity or FWHM. There does 
appear to be, however, a strong, significant drop in ionization state 
with distance from the nucleus, with [OIII]/[OII] $\propto r^{-1}$ 
[Fig.~\ref{fig1}(\textit{d})]. A proper interpretation of this trend 
is, nevertheless, quite complicated. Here, we can only note that the 
radial trend is evidence for a nuclear ionizing field in the NLR.

\subsection{Profile Comparisons}

  A hallmark of two-component models is that the line spectrum of each 
component is radically different. In the case of shocks, most of the flux
in high ionization lines is produced in the kinematically quiescent 
precursor, while the strongly disturbed postshock cooling region 
generates a low-ionization spectrum. Thus, it is reasonable to expect a 
shock-excited spectrum to show significantly different line kinematics 
between low-ionization lines, like [NII]$\lambda 6584$ and high ionization 
lines like [OIII]. This is applicable to the MB/IB scenario as well and 
can be used to set constraints on the level of co-spatiality of the two 
components. 

 To look for profile differences, we compared the [OIII] line profile to 
the profiles of a number of lines of different ionization species, after 
correcting for the wavelength shift and instrumental broadening. Only 
very minor differences were found, even in areas with strong signs of 
jet-gas interaction. We conclude that strong shocks are unlikely and 
components of different optical depths share a common velocity field.
   
\subsection{Estimates of Ionizing Luminosity}

  We can test for nuclear photoionization by comparing the UV ionizing flux 
of the AGN with the total emission line luminosity. Since the direct UV
flux of the AGN is obscured, we use the FIR luminosity as a surrogate for 
the dust-reprocessed UV luminosity. From IRAS FSC measurements and estimates
of geometrical and covering factors [see \cite{paper2}], we derive
$L_{UV}\sim L_{FIR}\sim 10^{43.5}\textrm{ erg s}^{-1}$, which is 
approximately equal to the total emission line luminosity, taken to be 
about $10\times L_{5007}$. Clearly, an ionizing field from the AGN 
is sufficient to power the observed line emission in Mkn 78 and there 
is no need for any additional source of ionizing photons, such as fast shocks.
  
 To conclude, all the evidence suggests that the principal ionizing source
of Mkn 78, and possibly most Seyferts, is the central AGN, coupled with
a realistic multi-component ionized gas distribution. The role of ionizing
shocks are negligible. This is borne out in detail in the dynamical analysis
of \cite{paper3}.


\begin{thebibliography}

\bibitem[Allen (2004)]{allen04}
Allen, M.G. 2004, private communication

\bibitem[Binette, Wilson, \& Storchi-Bergmann (1996)]{binette96}
Binette, L., Wilson, A.S., \& Storchi-Bergmann, T. 1996, A\&A, 312, 365

\bibitem[Baldwin, Phillips, \& Terlevich (1981)]{baldwin81}
Baldwin, J.A., Phillips, M.M., \& Terlevich, R. 1981, PASP, 93, 5

\bibitem[Davidson \& Netzer (1979)]{davidson79}
Davidson, K., \& Netzer, H. 1979, Rev. Mod. Phys. 51, 715

\bibitem[Dopita \& Sutherland (1996)]{dopita96}
Dopita, M.A., \& Sutherland, R.S. 1996, ApJS, 102, 161 

\bibitem[Ferland (1996)]{ferland96}
Ferland, G.J. 1996,  {\it Hazy, a Brief Introduction to Cloudy}, 
University of Kentucky Department of Physics and Astronomy Internal Report.

\bibitem[Robinson et al. (2000)]{robinson00}
Robinson, T.G., Tadhunter, C.N., Axon, D.J., \& Robinson, A. 2000, 
MNRAS, 317, 922

\bibitem[Viegas-Aldrovandi \& Contini (1989)]{viegas89}
Viegas-Aldrovandi, S.M., \& Contini, M. 1989, ApJ, 339, 689

\bibitem[Whittle (1992)]{whittle92}
Whittle, M. ApJ, 387, 109 

\bibitem[Whittle \& Wilson (2004)]{paper1}
Whittle, M., \& Wilson, A.S. 2004, AJ, 127, 606 (Paper I)

\bibitem[Whittle et al. (2004)a]{paper2}
Whittle, M., Rosario, D.J., Silverman, J.D., Nelson, C.H., \& Wilson, A.S.
2004a, AJ, submitted (Paper II).

\bibitem[Whittle et al. (2004)b]{paper3}
Whittle, M., Silverman, J.D., Rosario, D.J., Nelson, C.H., \& Wilson, A.S.
2004b, AJ, submitted (Paper III).

\end{thebibliography}
\end{document}